\documentclass{article}

\usepackage{PRIMEarxiv}

\usepackage[utf8]{inputenc} 
\usepackage[T1]{fontenc}    
\usepackage{hyperref}       
\usepackage{url}            
\usepackage{booktabs}       
\usepackage{amsfonts}       
\usepackage{nicefrac}       
\usepackage{microtype}      
\usepackage{lipsum}
\usepackage{fancyhdr}       
\usepackage{graphicx}       
\usepackage{multirow}
\usepackage{makecell}
\usepackage{amsmath}
\usepackage[section]{placeins}
\usepackage{array, multirow, boldline}
\title{Does spatial information improve influenza forecasting?
}

\author{
  Gabrielle Thivierge \\
  Department of Statistics and Data Science \\
  Carnegie Mellon University \\
  Pittsburgh, PA \\
  \texttt{gthivier@andrew.cmu.edu} \\
   \And
  Aaron Rumack \\
  Machine Learning Department \\
  Carnegie Mellon University \\
  Pittsburgh, PA\\
  \texttt{rumackaaron@gmail.com} \\
    \And
  F. William Townes \\
  Department of Statistics and Data Science \\
  Carnegie Mellon University \\
  Pittsburgh, PA\\
  \texttt{ftownes@andrew.cmu.edu} \\
}

\begin{document}
\maketitle

\begin{abstract}

Seasonal influenza forecasting is critical for public health and individual decision making. We investigate whether the inclusion of data about influenza activity in neighboring states can improve point predictions and distribution forecasting of influenza-like illness (ILI) in each US state using statistical regression models. Using CDC FluView ILI data from 2010-2019, we forecast weekly ILI in each US state with quantile, linear, and Poisson autoregressive models fit using different combinations of ILI data from the target state, neighboring states, and US weighted average. Scoring with root mean squared error and weighted interval score indicated that the variants including neighbors and/or the US average showed slightly higher accuracy than models fit only using lagged ILI in the target state, on average. Additionally, the improvement in performance when including neighbors was similar to the improvement when including the US average instead, suggesting the proximity of the neighboring states is not the driver of the slight increase in accuracy.
\end{abstract}

\section{Introduction}

An estimated 8\% of the US population is infected with seasonal influenza each year \cite{Tokars2018SeasonalStates}. Influenza case levels have a typical seasonal pattern, but it is not constant season-to-season or state-to-state. The ability to more accurately predict when and where influenza cases will increase can aid in public health decision making, resource allocation, and vaccine distribution \cite{Biggerstaff2016ResultsChallenge}.

Historical research on influenza forecasting has used influenza-like illness (ILI) as its target. A patient is classified as having an influenza-like illness if they have a fever of at least 100 F, and a cough and/or sore throat \cite{CentersforDiseaseControlandPrevention2024KeyFlu}. A positive influenza test is not needed for a case to be classified as ILI.

ILI information in the US is collected by the U.S. Outpatient Influenza-like Illness Surveillance Network (ILINet). This network consists of outpatient healthcare providers who report weekly patient visits for ILI, as well as the total number of  patient visits for any reason. Thus, ILI values can either be a count or the percentage of all patient visits that are due to ILI. ILINet providers can be found in all 50 states, Washington DC, Puerto Rico, and the US Virgin Islands, providing a broad snapshot of influenza activity in the United States \cite{CentersforDiseaseControlandPrevention2024U.S.Methods}.

There are many possible modeling approaches used in epidemiological forecasting, which can broadly be broken down into mechanistic and statistical models and may include spatial, temporal, or spatiotemporal information \cite{Chretien2014InfluenzaReview}. While mechanistic models are useful for many tasks, particularly scenario analysis, there is some evidence that statistical models may demonstrate superior performance in forecasting tasks: Banholzer et al. (2023) found that in COVID-19 forecasts, statistical models performed at least as well as mechanistic models and better captured volatility \cite{Banholzer2023AModels}. 

The US COVID-19 ForecastHub includes examples of many mechanistic and statistical forecasters, but few of these incorporate spatial information  \cite{Cramer2022TheDataset}. The increased computational complexity from incorporating spatial data could be a reason these models often exclude this information. Of those that do incorporate spatial information, most are mechanistic models, such as the MOBS-GLEAM model, an agent-based model that incorporates human mobility data to predict the spread of disease \cite{Balcan2010ModelingModel, Tan2022StatisticalRegions}. An exception is Osthus and Moran's Dante model (2021), a Bayesian hierarchical model consisting of a state-level submodel and a national-level submodel, which models patterns that are common across states \cite{Osthus2021MultiscaleForecasting}. This model won the 2018/19 FluSight challenge, which suggests that spatial information can be beneficial for ILI forecasting.
 
With this context, we aim to investigate the utility of directly including spatial information in simple autoregressive models. Specifically, at the US state level, we wish to determine if information about surrounding states improves model performance within a state. In this analysis, we explore the effect of this data on linear, quantile, and Poisson autoregressive models fit at the state level that may or may not use neighbor state data or the US average ILI. We do not focus on performance between autoregressive model classes; instead, we focus on the difference in spatial variants within classes. Here, we are most interested in the impact of the spatial information on performance rather than a comparison of the accuracy of Poisson, quantile, and linear regression models. 


\section{Methods}
\label{sec:methods}

Using influenza-like illness (ILI) data, we fit linear, Poisson, and quantile autoregressive models. We will refer to linear, quantile, and Poisson regressions as "Model Classes" (Table \ref{tab:model-classes}), which differ in their method of parameter estimation. To explore the role of spatial information, within each of these three classes, we used five different subsets of covariates. We refer to these as the "Model Variants" (section \ref{model_variants}). With five model variants and three model classes, we have 15 regression models. In addition to these 15, we computed a last-value-carried-forward time series, for a total of 16 models, each of which were then applied for each state and time point as described in section \ref{fitting}.


\begin{table}[h]
\centering
\caption{Lists of model classes and variants}
\label{tab:model-classes}
\begin{tabular}{@{}clllc@{}}
\cmidrule(r){1-1} \cmidrule(l){5-5}
\textbf{Model Classes}     &  &  &  & \textbf{Model Variants} \\ \cmidrule(r){1-1} \cmidrule(l){5-5} 
Linear (L)                    &  &  &  & Isolated                \\
Poisson (Po)                   &  &  &  & Neighbors               \\
Quantile (Q)                  &  &  &  & Isolated + US Average   \\
Last Value Carried Forward &  &  &  & Neighbors + US Average  \\ \cmidrule(r){1-1} \cmidrule(l){5-5} 
\end{tabular}
\end{table}

\subsection{Data}
The ILI data used in the analysis come from CDC FluView \cite{CDCInfluenzaDivision2024WeeklyReport}, accessed via the Delphi Epidata API \cite{epidatr}. Observations represent the state-level and US weighted average ILI count and percentage each week, from October 2010 through May of 2019. We included only the final data revision at each date. Seasons after 2019 were excluded to avoid bias in ILI due to the COVID-19 pandemic. Because reporting is less consistent in the off-season, only observations for epiweeks 40-52 (October-December) and 1-18 (January - May) were used for model fitting. All models used the 156 in-season weeks across the 2010-11, 2011-12, 2012-13, 2013-14, and 2014-15 seasons for training.

\subsection{Model fitting}\label{fitting}

For each model class and variant, beginning with the 2015-16 season, we estimated parameters at each successive time point to account for nonstationarity. That is, all models were fit using the 156 in-season training data points and test season data up to the date of prediction. Thus, models were fit starting with week 253 through week 448, April 2019, which constitutes 124 weeks after excluding out-of-season points. The iterative re-fitting reflects the information gained as new measurements are taken each week. This also means that each prediction comes from a different model, fit on slightly different data.

For every model class and variant except the geo-pooled model, each of the 50 US states had a unique autoregressive model fit at each time point. The geo-pooled model assumes dynamics are constant across locations, and so each state is predicted using the same model.

In the linear and quantile classes, the outcome variable was \% ILI, and the predictor variables were also \% ILI. For the Poisson class, the outcome variable was the weekly ILI count rather than percentage. The log of total patient visits were used as an offset. Predictor variables were also on the percentage scale as in the linear and quantile classes.

\subsection{Baseline method}

A last-value-carried-forward (LVCF) model was used as a baseline for the analysis. For each state $\ell$, the predicted ILI at time $t+2$ was simply the ILI in state $\ell$ at time $t$:

\begin{align*}
     E\left[ I_{\ell, t+2} \right] = I_{\ell, t}
\end{align*}

\subsection{Basic linear regression model}\label{linear}

Let $S(\ell)$ be the set of states that border state $\ell$. Then let $I_{s, t}$ be the \% ILI in state $s \in S(\ell)$ at time $t$.  Additionally, let $D_{S, k}$ be an indicator representing whether cases are increasing, decreasing, or flat in covariate state $s$ from $t - k - 1$ to $t-k$ according to

\begin{align*}
D_{s, t, k} = \left\{ \begin{array}{ll} 1 \qquad I_{s, t-k} - I_{s, t - k-1)} > \epsilon \\[1em] -1 \qquad I_{s, t-k} - I_{s, t - k-1)} < -\epsilon \\[1em] 0 \qquad \left| I_{s, t-k} - I_{s, t - k -1)} \right| \leq \epsilon \end{array} \right.      
\end{align*}
for a small tolerance $\epsilon = 0.05$. Because ILI is unlikely to be exactly equal week-over-week, values within $\epsilon$ of one another are considered 'flat' for this analysis. We chose to use this indicator rather than the exact values because ILI values are not necessarily comparable between states \cite{Rumack2023CorrectingIndicators}. Here, we use $k \in \{0,1\}$ to represent the two lagged trend indicators in the neighboring states, i.e., the changes from week $t-2$ to $t-1$ and $t-1$ to $t$.



Then, the linear regression model to predict \% ILI in target state $\ell$ at time $t+2$ is given by

\begin{align*}
    E\left[ I_{\ell, t+2} \right] = \sum_{k = 0}^2 I_{\ell, t - k} \beta^{(L)}_{\ell, t, k}  + \sum_{s \in S(\ell)} \sum_{k = 0}^1  D_{s, t, k} {\gamma}^{(L)}_{s, t, k} + \mu^{(L)}_{\ell, t}
\end{align*}

where $\beta^{(L)}_{\ell, t, k}$ is the regression coefficient corresponding to the lagged \%ILI in target state $\ell$, at time $t - k$, fit on data for time $t' = 3$ to $t$. The superscript (L) refers to the model class: linear (L), quantile (Q), and Poisson (Po). For this analysis, we used $k \in \{0,1,2\}$, which represent three lagged \%ILI values in the target state.  Similarly, $\gamma^{(L)}_{s, t, k}$ is the linear regression coefficient for the indicator variable representing change in \%ILI for neighboring state $s$ from time $t-k-1$ to $t-k$, fit on data for time $t' = 3$ to $t$. The term $\mu_{\ell, t}$ represents the intercept term for the model fit using data from $t' = 3$ to $t$ to predict \%ILI in target state $\ell$. Note that the time subscript on the coefficients means that each prediction is made by a different regression model, though the models for times $t$ and time $t+1$ are very similar, since there is only a single observation difference in their training data. 

We also fit quantile and Poisson autoregressive models, which are specified in section \ref{addtl_models}.

\subsection{Model variants}\label{model_variants}

\begin{figure}[htp]
    \centering
    \includegraphics[width=12cm]{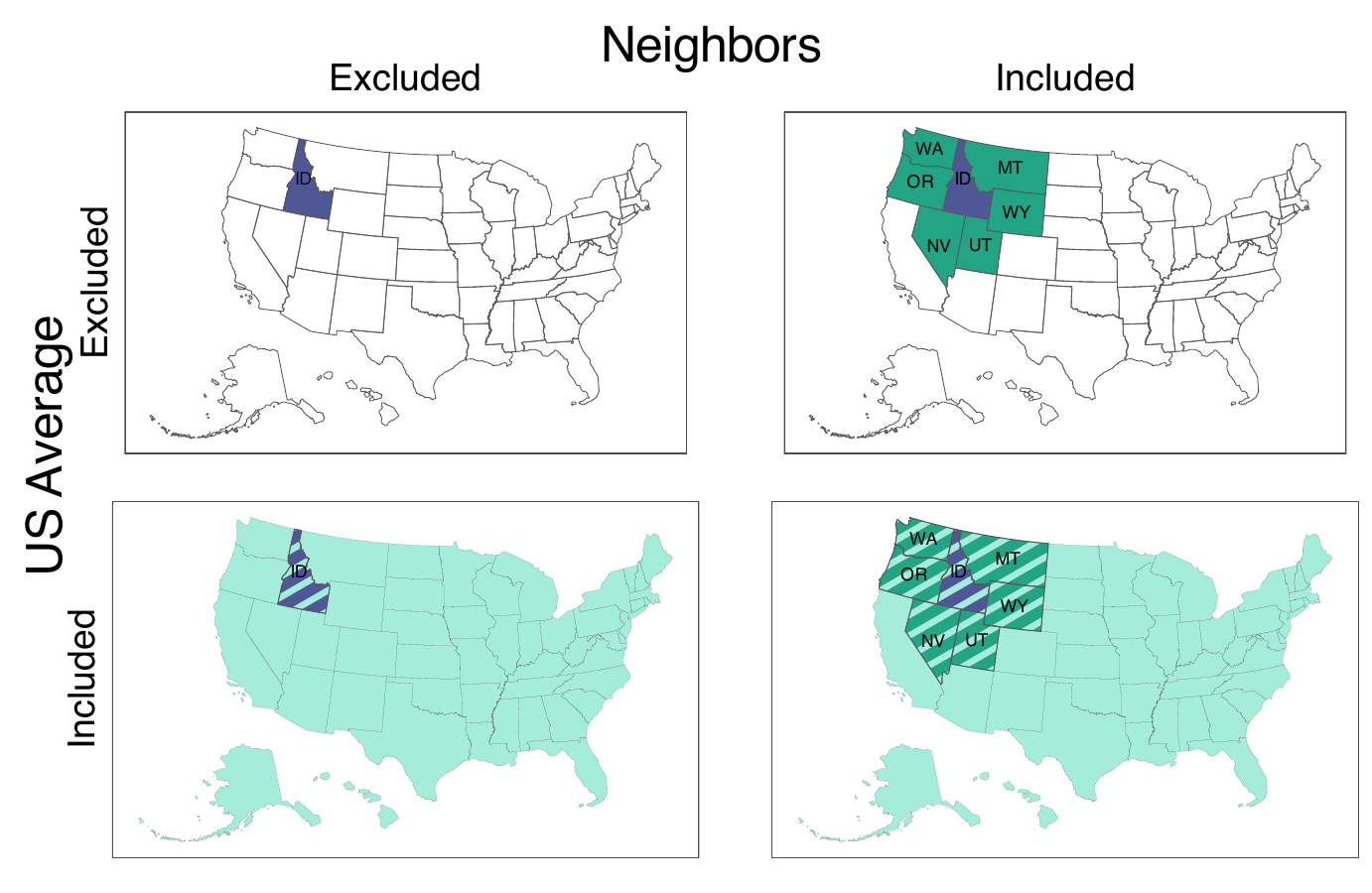}
    \caption{Model variants use different combinations of Neighbors and US Average as additional covariates. These maps show an example of four model variants when Idaho is the target state. Clockwise from top left, the maps correspond to the Isolated, Neighbors, Neighbors + US Average, and Isolated + US Average model variants. Solid colors indicate the inclusion of the lags or lag indicators for a given state, while stripes indicate that the state information is also incorporated into the weighted US Average \%ILI in addition to its lag terms.}
    \label{fig:model_vars}
\end{figure}

To explore the impact of spatial information on predictive performance, we fit several variations on the basic linear autoregressive model for each state, at each time point. We applied every model variant to each of the three model classes. A visual demonstration of the Isolated, Neighbors, Isolated + US Average, and Neighbors + US Average models, using Idaho as the example target state, is provided in Figure \ref{fig:model_vars}. Each state has its own set of models for all variants except geo-pooled, which is fit using all states together.

\textbf{Isolated}: for target state $\ell$ at time $t$, include only terms for the target state's own lags, so that
\begin{align*}
      E\left[ I_{\ell, t+2} \right] = \sum_{k = 0}^2 I_{\ell, t - k} \beta_{\ell, t, k} 
\end{align*}

\textbf{Neighbors}: for target state $\ell$ use $I_{\ell, t - k}$ where $k \in \{0,1,2\}$ and $D_{s, t, k}$ where $k \in \{0,1\}$ for all $s \in S(\ell)$, the states bordering state $\ell$. This formulation is specified as the basic regression in section \ref{linear}.

\begin{align*}
    E\left[ I_{\ell, t+2} \right] = \sum_{k=0}^2 I_{\ell, t - k} \beta_{\ell, t, k}  + \sum_{s \in S(\ell)} \sum_{k=0}^1  D_{s, t, k} {\gamma}_{s, t, k} + \mu_{\ell, t}
\end{align*}

\textbf{Isolated + US Average}: for target state $\ell$ use $I_{\ell, t}$ and $D_{US, t, k}$ where $D_{US, t, k}$ represents whether cases are increasing, decreasing, or flat in the weighted US average ILI over the previous two one-week periods, in the same manner as the neighboring states.

\begin{align*}
    E\left[ I_{\ell, t+2} \right] = \sum_{k=0}^2 I_{\ell, t - k} \beta_{\ell, t, k}  + \sum_{k=0}^1  D_{US, t, k} {\theta}_{t, k} + \mu_{\ell, t}
\end{align*}

\textbf{Neighbors + US average}: US Average: for target state $\ell$ use the Neighbors and US average terms as described previously, so that

\begin{align*}
    E\left[ I_{\ell, t+2} \right] = \sum_{k=0}^2 I_{\ell, t - k} \beta_{\ell, t, k}  + \sum_{s \in S(\ell)} \sum_{k=0}^1  D_{s, t, k} {\gamma}_{s, t, k} + \sum_{k=0}^1  D_{US, t, k} {\theta}_{t, k} + \mu_{\ell, t}
\end{align*}

For the purposes of the "Neighbors" model variants, Alaska was assumed to `border' Washington, and Hawaii was assumed to `border' California, Washington, and Oregon.

\textbf{Geo-pooled}\label{gp}: We fit one model using all states together at each time point, representing an assumption that the autoregressive dynamics are constant across states,  i.e., use $I_{\ell, t - k}\ \forall\ \ell \:$ in model fitting. Thus the regression coefficients are no longer location-specific. 

\begin{align*}
    E\left[ I_{\ell, t+2} \right] = \sum_{k=0}^2 I_{\ell, t - j} \beta_{t, k} + \mu_{t} 
\end{align*}

where $\beta$ is estimated by fitting a linear model on all states, i.e., minimizing the loss function
\begin{align*}
    \sum_{t' = 3}^t \sum_{\ell = 1}^{50} \left( I_{\ell, t'+2} -  E\left[ I_{\ell, t'+2} \right]\right)^2 
\end{align*}

\subsection{Additional regression models}{\label{addtl_models}}

The quantile regression to predict the median \%ILI in state $\ell$ at time $t+2$ is given by

\begin{align*}
    Median(I_{\ell, t+2}) = \sum_{k=0}^2 I_{\ell, t-k} \beta^{(Q)}_{\ell, t}  + 
    \sum_{s \in S(\ell)} \sum_{k=0}^1  D_{s, t, k} {\gamma}^{(Q)}_{s, t, k} + \mu^{(Q)}_{\ell, t}
\end{align*}

 In this analysis, we only computed the median quantile, as it was the only one needed for our method of creating confidence intervals.  

The Poisson regression requires some additional notation to account for the use of count data rather than \%ILI as the outcome. Let $C_{\ell, t}$ be the count of ILI visits in target state $\ell$ at time $t$, and let 
$P_{\ell, t}$ be the number of providers reporting case data in state $\ell$ at time $t$. Also, let $V_{\ell, t}$ be the total number of reported patient visits in state $\ell$ at time $t$. Note that $I_{\ell, t} = \frac{C_{\ell,t}}{V_\ell,t}$.

Then, given the previous notation and 

\begin{align*}
     &M_{s, t, k} = \left\{ \begin{array}{ll} 1 \qquad \log(I_{s, t-k}) - \log(I_{s, t - k- 1}) > \epsilon \\[1em] -1 \qquad \log(I_{s, t-k}) - \log(I_{s,t - k- 1}) < -\epsilon \\[1em] 0 \qquad \left| \log(I_{s, t-k}) - \log(I_{s, t - k- 1}) \right| \leq \epsilon
     \end{array} \right.
\end{align*}
the Poisson regression to predict the number of ILI cases in location $\ell$ at time $t+2$ is given by

\begin{align*}
    \log{(E\left[C_{\ell, t+2}\right])} &= \log(c + V_{\ell, t+2}) +  \log{(c + P_{\ell, t+2})} \phi_{\ell,t+2} + \sum_{j} \log(I_{\ell, t-j}+c) \beta^{(Po)}_{\ell, t}  + \\ &\sum_{s \in S(\ell)} \sum_{k}  M_{s, t, k} {\gamma}^{(Po)}_{s, t, k} + \mu^{(Po)}_{\ell, t}
\end{align*}

where 
\begin{align*}
    c = \frac{1}{2} \underset{s,t}{\min}(I_{s,t})  \approx 0.0022, \: t \in {1...T}, \: s \in \{\text{US States}\}
\end{align*}

is a small constant added to avoid division by 0 in the case of missing data, and $\phi_{\ell,t+2}$ is the regression coefficient corresponding to the log of the number of patient visits in state $\ell$ at time $t+2$. 

\subsection{Prediction and performance}

\subsubsection{Forecasting}

For all but the geo-pooled model, each prediction comes from a model fit unique to the state and time point. For the geo-pooled model, each prediction comes from a model fit unique to the time point, but not unique to the state. At time $t$, we fit each model on all data for all time points $t' < t$, and used it to predict the value at $t+2$. 

After obtaining point predictions, we created 50, 80, and 95 \% intervals using the quantile tracker as described in \cite{Angelopoulos2023ConformalPrediction}. This conformal prediction method is nonparametric (i.e., it makes no distributional assumptions) and creates intervals using the sequence of point predictions. 

\subsubsection{Performance}
We focused on the accuracy of point predictions and interval forecasts when averaged over states for each time point as well as when averaged over time for each state.

To  measure the performance of the point predictions, we employed the root mean squared error (rMSE) of the predictions. Let $T$ be the total number of weeks of data in the train and test sets combined, and let $t^*$ be the first week of the test data set. We evaluated this metric by averaging over all test time points for each state, such that 

\begin{align*}
    rMSE_{\ell} = \left[\frac{1}{T-t^*} \sum_{t = t^*}^T (\hat{I}_{\ell,t} - I_{\ell, t})^2\right]^{1/2}
\end{align*}

 where $\hat{I}_{\ell,t}$ represents the predicted value of $E[I_{\ell, t}]$, or by averaging over all 50 states for a given time point, such that

\begin{align*}
    rMSE_{t} = \left[\frac{1}{50} \sum_{\ell = 1}^{50} (\hat{I}_{\ell,t} - I_{\ell, t})^2\right]^{1/2}
\end{align*}

To measure performance of the distribution, we used the weighted interval score (WIS) as described in \cite{Bracher2021EvaluatingFormat}. Here, we suppress the location subscripts for clarity. The weighted interval score is, as the name suggests, a weighted average of interval scores, where the interval score for an $\alpha \%$ prediction interval $\left[ \hat{y}_{\alpha/2, t}, \: \hat{y}_{1- \alpha/2
, t}\right]$ and the true value of $y_t$ is given by

\begin{equation*}
    \text{IS}_{\alpha, t} = (\hat{y}_{1-\alpha/2, t} - \hat{y}_{\alpha/2, t}) + \frac{2}{\alpha}(\hat{y}_{\alpha/2, t} - y_t) * \mathbb{I}(\hat{y}_{\alpha/2, t} < y) + \frac{2}{\alpha}(y_t - \hat{y}_{1-\alpha/2, t}) * \mathbb{I}(y_t > \hat{y}_{1-\alpha/2, t})
\end{equation*}

Then, the WIS for a state $\ell$ at time $t$ is given by

\begin{equation*}
    WIS_{\ell, t} = \frac{1}{K + 1/2} \left( \frac{w_0}{2} | y - m| +  \sum_{k=1}^{K} w_k \cdot IS_{\alpha_k, t} \right)
\end{equation*}

where $w_k = \frac{\alpha}{2}$, $w_0 = 0.5$, and $m$ is the estimated median ILI. In this analysis, we used values $\alpha$ = 0.5, 0.8, and 0.95.

As with the rMSE, the WIS was computed for each prediction and averaged for each model variant for each state over time points in the test set, given by 

\begin{align*}
    WIS_{\ell} = \frac{1}{T-t^*} \sum_{t = t^*}^T WIS_{\ell, t}
\end{align*}

and averaged for each model variant across all 50 states at a given time point, given by

\begin{align*}
    WIS_{t} = \frac{1}{50} \sum_{\ell = 1}^{50} WIS_{\ell, t}
\end{align*}

\subsection{Implementation}

We performed all analyses using R version 4.3.0 \cite{rstudio}. Quantile regression models were fit using the \texttt{quantreg} package (version 5.96) \cite{quantreg}, and linear models were fit using \texttt{limSolve} (version 1.5.7.1) \cite{limsolve, limsolve2}. Interval scores were computed via the \texttt{scoringutils} package (version 1.1.0) \cite{scoringutils}. 

\section{Results}
\label{sec:results}

Forecasts were evaluated by comparing rMSE and WIS between model variants within classes, as the focus of this work is on the relative benefit of spatial information rather than searching for a global forecasting strategy. Lower values of rMSE and WIS indicate reduced forecasting error. 


\subsection{Spatial information improves forecast accuracy across seasons}

\begin{figure}[htp]
    \centering
    \includegraphics[width=16cm]{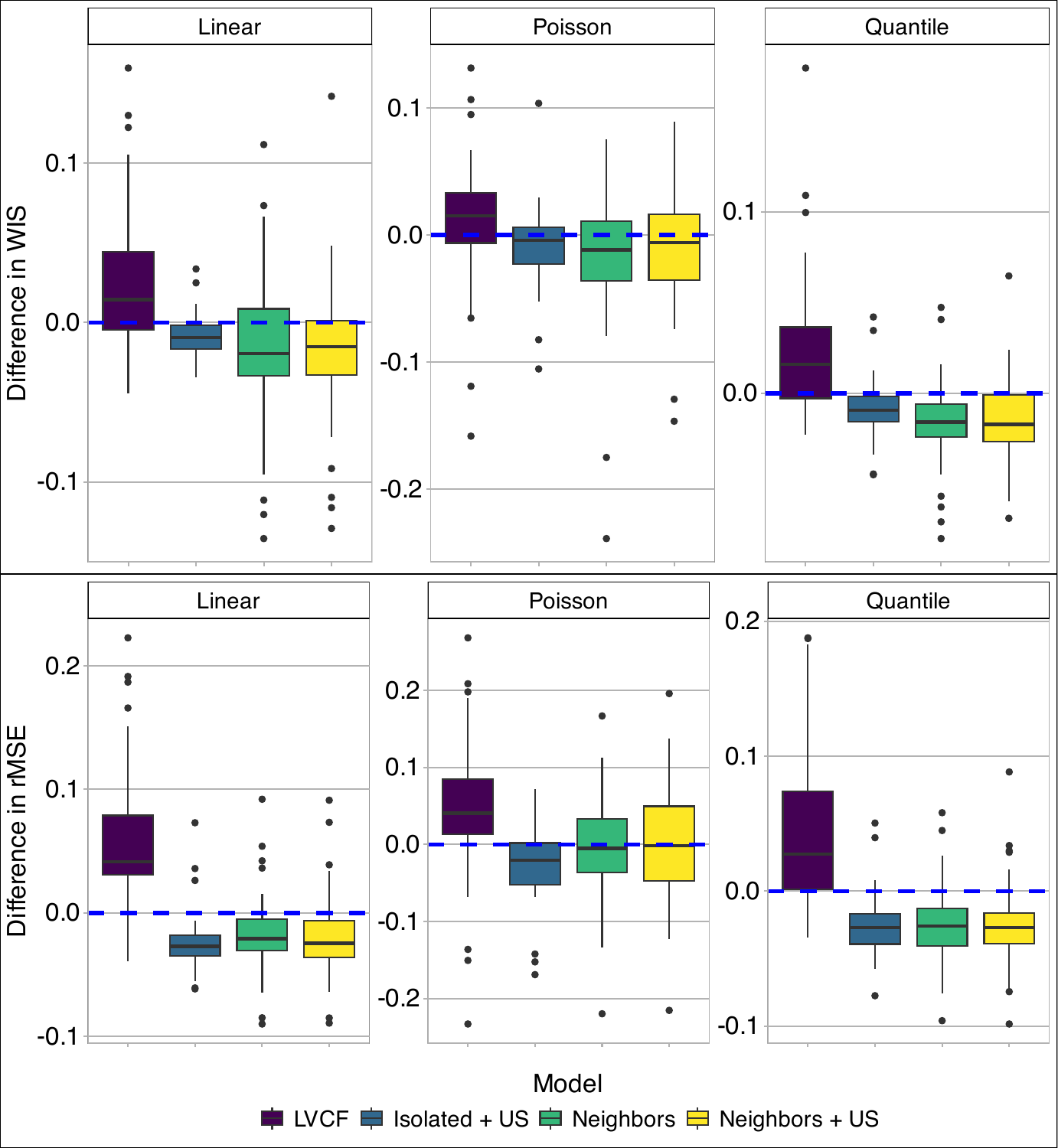}
    \caption{Spatial variants improve accuracy relative to the Isolated model for quantile and linear regressions, but not Poisson. rMSE and WIS values were averaged within each state across all seasons, and then subtracted from that of the state's Isolated model. Negative values indicate that the model variant reduced prediction error relative to the Isolated model. }
    \label{fig:isol_baseline_exgeo}
\end{figure}

When considering the last value carried forward (LVCF) model as the baseline and averaging over all time points for each state, only the Geo-pooled model demonstrated higher rMSE and WIS (Figure \ref{fig:lvcf_baseline}). All autoregressive models outperformed the last value carried forward (LVCF) model on average, when averaging rMSE and WIS within states over seasons (Figure \ref{fig:isol_baseline_exgeo}). For the linear and quantile classes, including spatial information (neighbors and/or US average) improved accuracy relative to the Isolated variant in terms of both point predictions and interval forecasts. For the Poisson class, including spatial information did not improve the accuracy of point predictions or distribution forecasts, on average. 

Across model classes, the variants that include the Neighbors data performed similarly to one another, but were not consistently more or less accurate than the Isolated + US Average model. The Neighbors models also demonstrate higher spread in the average WIS and rMSE difference in comparison to the Isolated + US Average model. This suggests that for some states, the Neighbors data had higher magnitude effect on performance than the US Average data. Figures \ref{fig:isol_baseline_wis_szn} and \ref{fig:isol_baseline_rmse_szn} demonstrate the difference in WIS and rMSE relative to the Isolated model broken out by season, respectively.

\subsection{Improvement in accuracy differs between seasons}

\begin{figure}[htp]
    \centering
    \includegraphics[width=16cm]{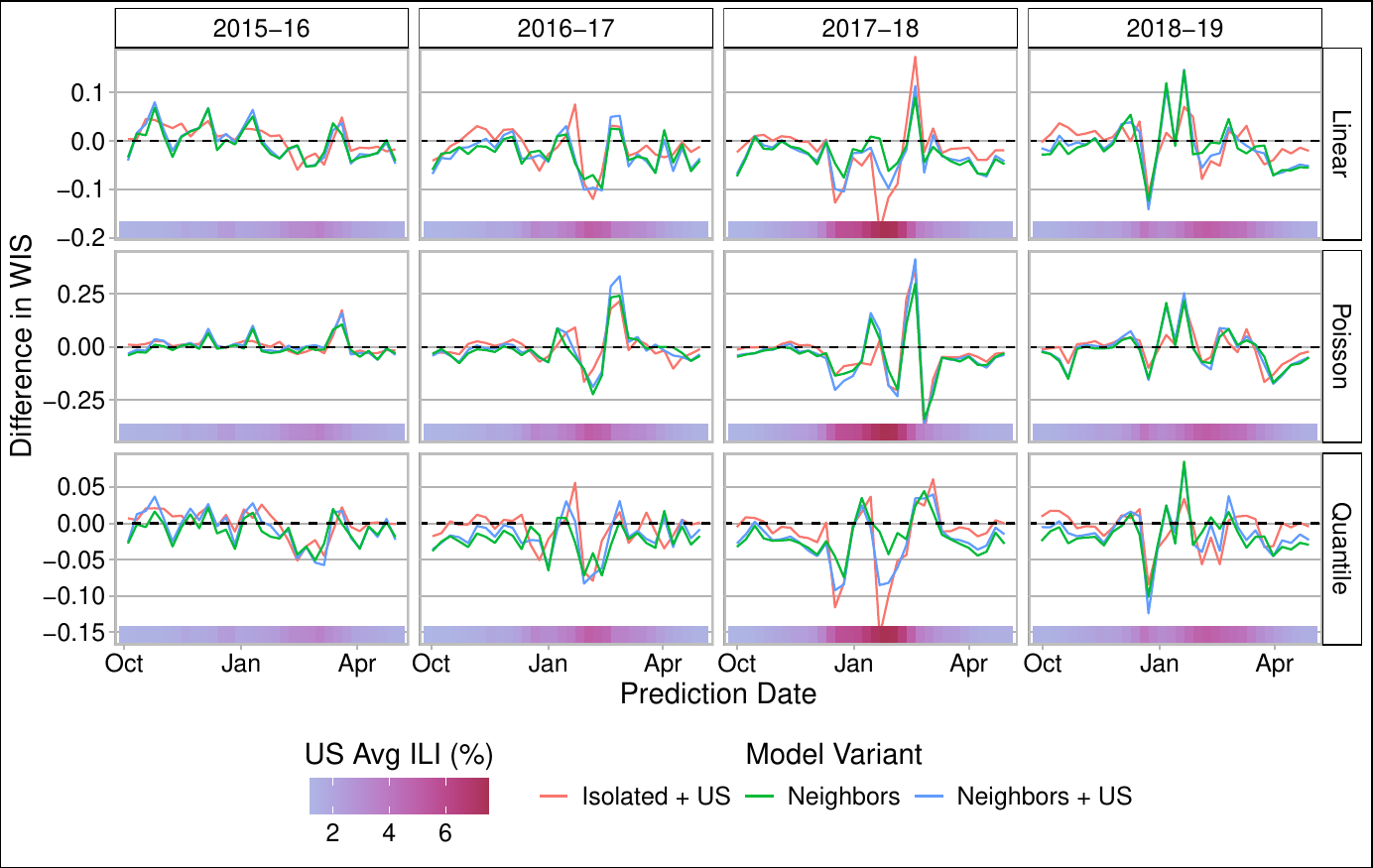}
    \caption{The effect of spatial information on interval prediction accuracy varies throughout the influenza season. At each time point, the WIS values for the predictions in each state were averaged and subtracted from the that of the Isolated model at the same time point. Negative values indicate that the model variant reduced WIS relative to the Isolated model.  The rug plots illustrate the US average ILI at each time point, for context regarding the seasonal influenza activity level.}
    \label{fig:isol_bl_timeseries_wis_split}
\end{figure}

\begin{figure}[htp]
    \centering
    \includegraphics[width=16cm]{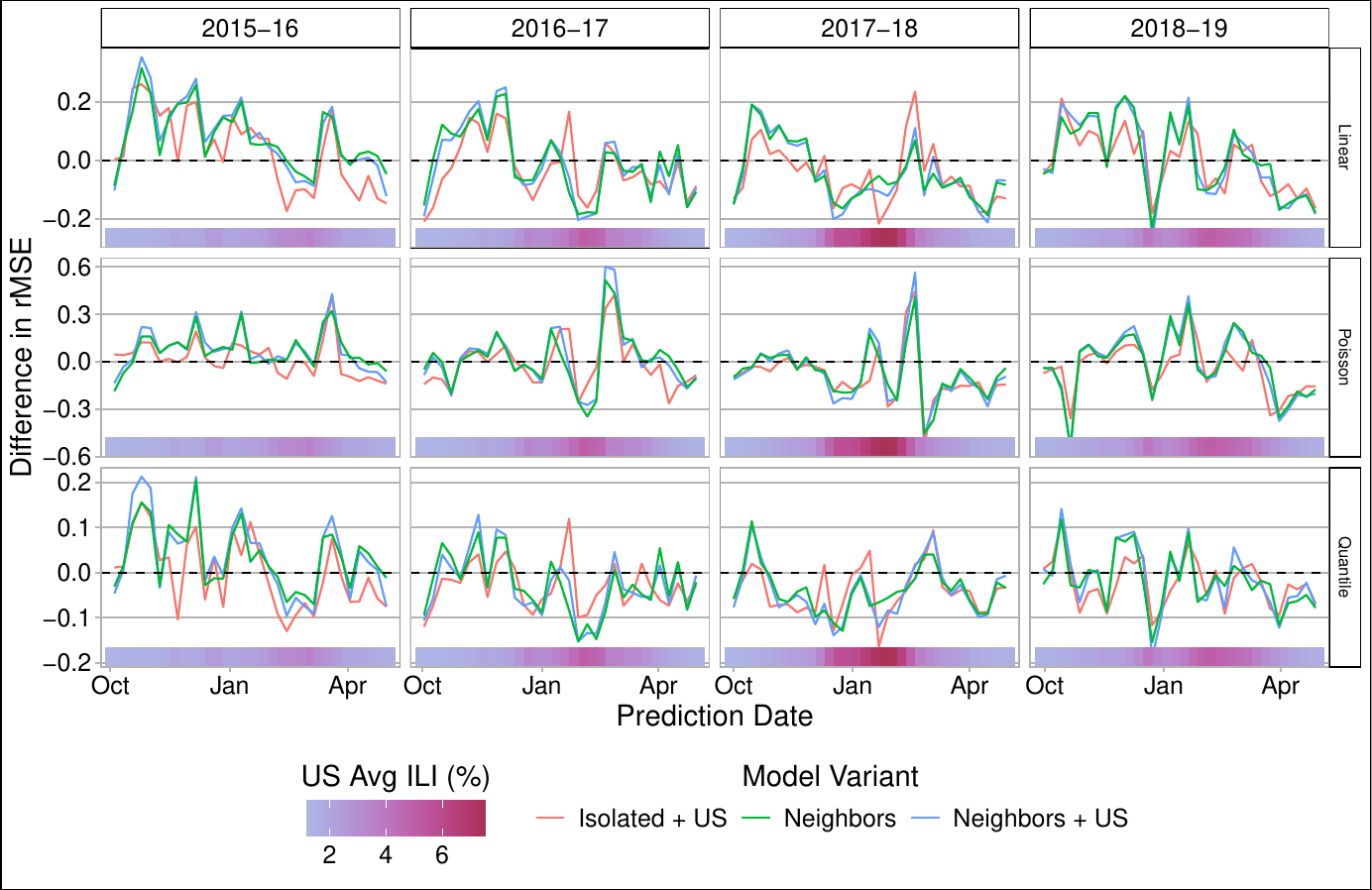}
    \caption{The effect of spatial information on point prediction accuracy varies throughout the influenza season. At each time point, the rMSE values for the predictions in each state were averaged and subtracted from the that of the Isolated model at the same time point. Negative values indicate that the model variant reduced rMSE relative to the Isolated model.  The rug plots illustrate the US average ILI at each time point, for context regarding the seasonal influenza activity level.}
    \label{fig:isol_bl_timeseries_rmse_split}
\end{figure}

The difference in accuracy relative to the Isolated model is not constant over time (Figures \ref{fig:isol_bl_timeseries_wis_split} and \ref{fig:isol_bl_timeseries_rmse_split}). There is clear within-season and between-season variability in the effect of spatial information on predictive performance. Interestingly, in the 2016-17 season, there is a sharp increase in model accuracy during the most intense part of the influenza season, followed by a sharp decrease in accuracy after the peak. In the 2017-18 and 2018-19 seasons, there is also an increase in accuracy when influenza activity intensifies, but this is not constant throughout the period of peak activity.  

\subsection{Accuracy improvement differs between states}

\begin{figure}[htp]
    \centering
    \includegraphics[width=\linewidth]{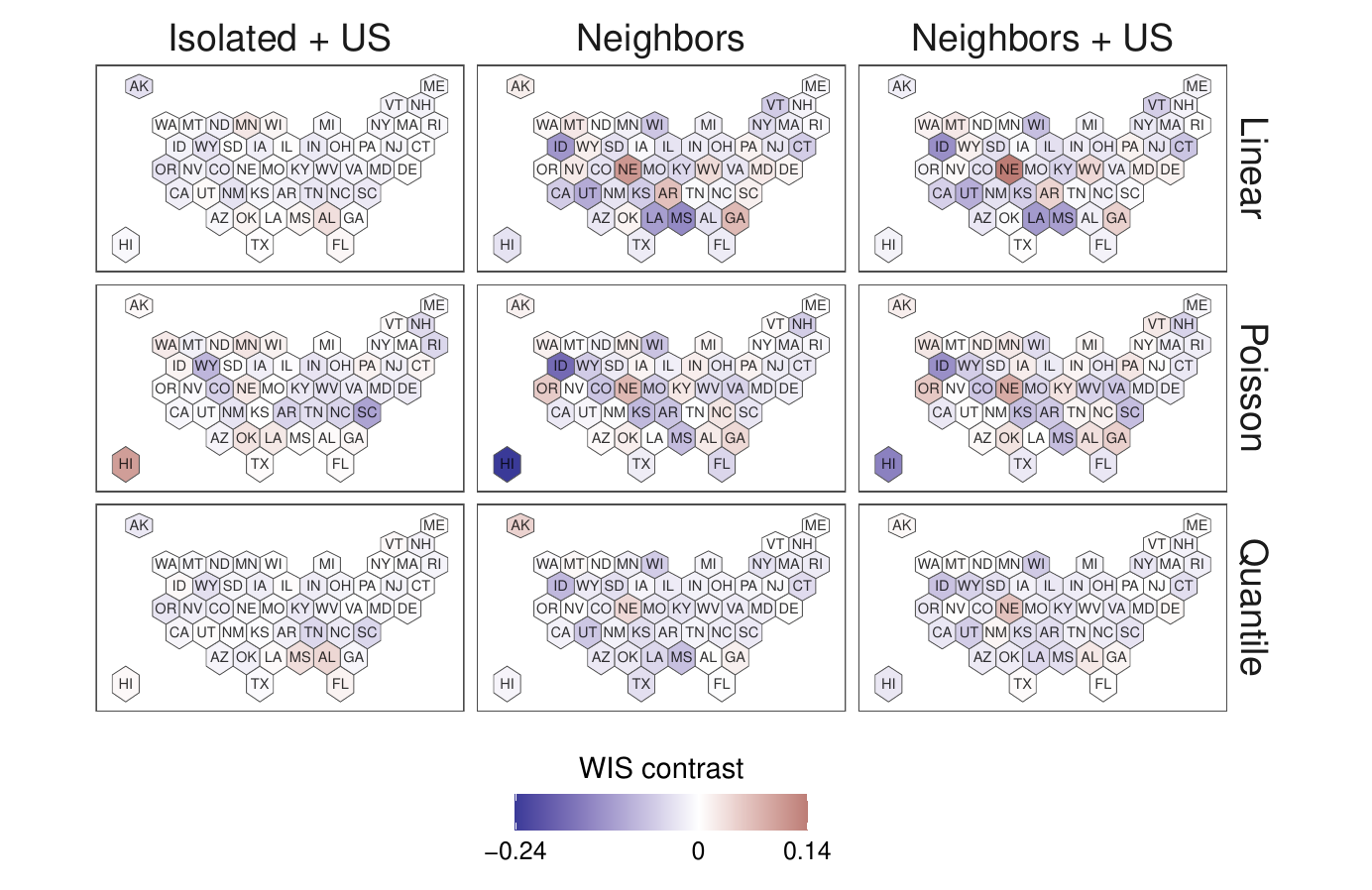}
    \caption{The difference in mean WIS relative to the Isolated model varies by state. At each time point, the WIS values for the predictions in each state were averaged and subtracted from the that of the Isolated model at the same time point. Negative values indicate that the model variant reduced WIS relative to the Isolated model.}
    \label{fig:isol_bl_map_wis}
\end{figure}

\begin{figure}[htp]
    \centering
    \includegraphics[width=\linewidth] {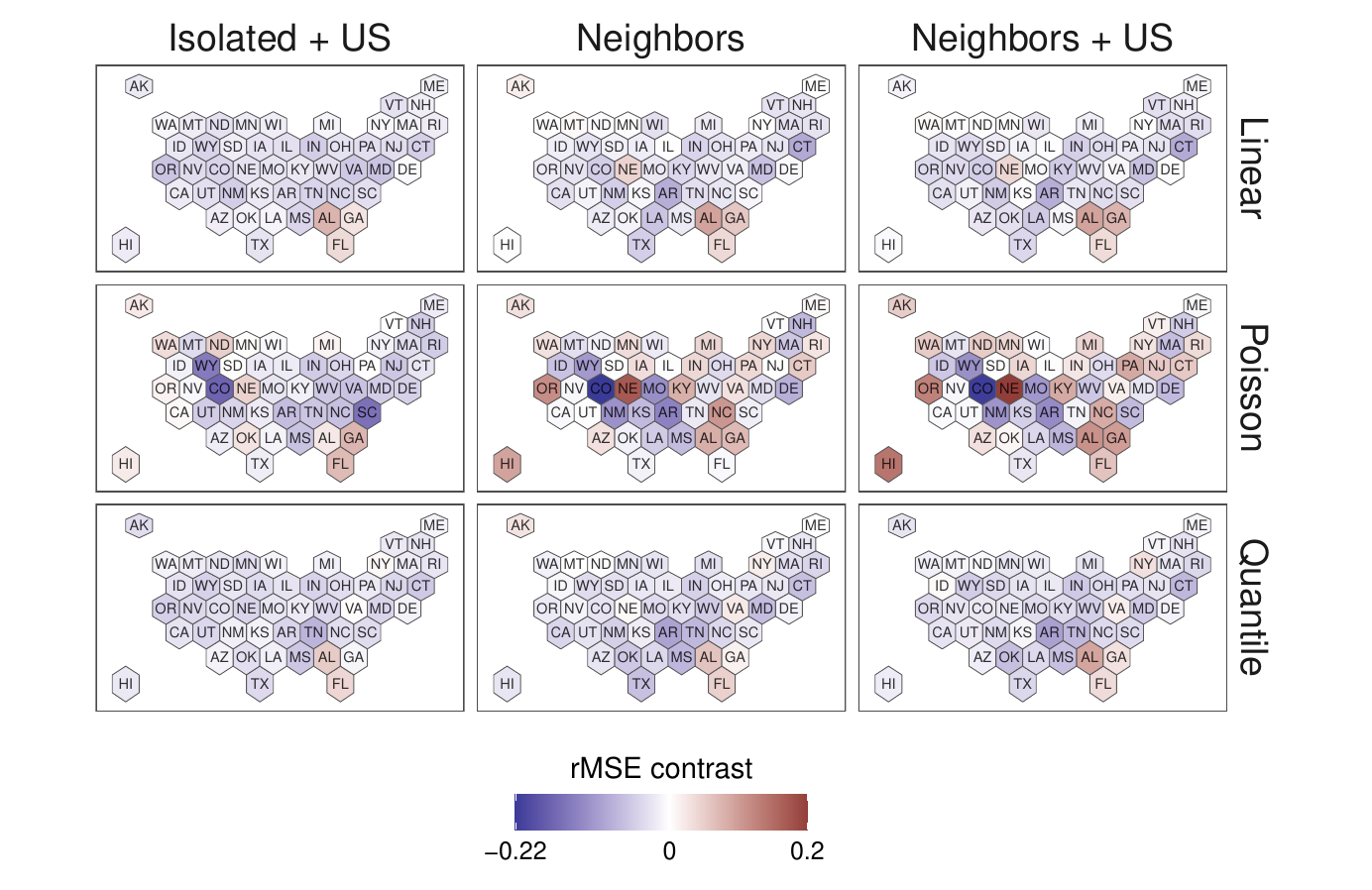}
    \caption{The difference in mean rMSE relative to the Isolated model varies by state. At each time point, the rMSE values for the predictions in each state were averaged and subtracted from the that of the Isolated model at the same time point. Negative values indicate that the model variant reduced rMSE relative to the Isolated model.}
    \label{fig:isol_bl_map_rmse}
\end{figure}

There is considerable variability of the effect of spatial information on performance between states (Figures \ref{fig:isol_bl_map_wis} and \ref{fig:isol_bl_map_rmse}). Certain states, such as Hawaii (HI) and Idaho (ID), see a consistent improvement in WIS when including spatial information, regardless of the use of a quantile, linear, or Poisson regression. Other states, such as Nebraska (NE), consistently demonstrate higher prediction error with the addition of the spatial data. These trends are also often consistent within model classes across seasons (Figures \ref{fig:isol_bl_map_wis_szn_lin}, \ref{fig:isol_bl_map_wis_szn_q}, \ref{fig:isol_bl_map_rmse_szn_lin}, and \ref{fig:isol_bl_map_rmse_szn_q}). There does not appear to be a regional trend in whether spatial information improves accuracy; there are states in all regions of the country whose models demonstrate better accuracy with spatial information.

It is important to note that a large difference in rMSE or WIS does not necessarily imply a difference of similar magnitude or direction in the other metric. For example, model variants for Hawaii that include neighbor data demonstrate more accurate interval forecasts, but they also generate less accurate point predictions. 

\newpage

\section{Discussion}

Including the US average ILI and/or data from neighboring states in autoregressive models for ILI forecasting improved point prediction and interval forecast accuracy relative to an Isolated model, on average, for linear and quantile regressions. However, this trend was not seen for Poisson models, though it was not immediately clear why. Additionally, we have shown that the improvement in model performance when including neighboring states is similar to the improvement when including the US average. This suggests that the activity in neighboring states is not necessarily useful due to proximity itself, but rather the information gained from regularizing toward the overall trend, similar to the Dante model \cite{Osthus2021MultiscaleForecasting}.

The coarse resolution of the spatial data may not be informative enough to improve regression performance consistently. Previous work, such as Charu et al. (2017) \cite{Charu2017HumanStates} found spatial patterns of influenza transmission at the city level. States likely do not have uniform distribution of influenza cases within their borders, and so the existence of a border between states does not guarantee that high influenza activity in one state necessarily affects its neighbors. For example, high ILI in Colorado might not affect Oklahoma a great deal, since their population centers are very separated.

There are further limitations to influenza-like illness as an outcome metric beyond the coarse spatial resolution. There is spatial and temporal heterogeneity in ILI reporting, and counts may not be comparable between states \cite{Rumack2023CorrectingIndicators}. Because it is a voluntary reporting network, the number of providers reporting in each state can change throughout and between influenza seasons. Providers may not be uniformly distributed throughout a state, and so certain areas may be over- or underrepresented in ILI counts, which could affect the utility of the neighbors data for certain states. This idea is supported by our finding that some states demonstrated consistent differences in model accuracy across model variants and seasons, either positive or negative, relative to the Isolated model when adding Neighbors data. 

Also, ILI does not uniquely capture influenza cases; because there is no requirement for a positive influenza test, other respiratory illnesses may be classified as ILI. With the arrival of COVID-19, recent ILI values are not comparable to pre-pandemic measurements. Measurements such as influenza hospitalizations collected from the CDC RESP-NET network could provide a better estimate of true influenza activity and would be of interest for further exploration in forecasting. 


Overall, our results suggest that at the state level, ILI forecasting accuracy is slightly improved by data from neighboring states; however, we hypothesize that spatial information at a different resolution could lead to a greater boost in forecasting accuracy. Further work at a finer spatial resolution and with alternative metrics such as hospitalizations or lab-confirmed cases might demonstrate a greater value of spatial information for influenza forecasting with autoregressive models. 

\section{Acknowledgements} 

This material is based upon work supported by the United States of America Department of Health and Human Services, Centers for Disease Control and Prevention, under award number NU38FT000005; and contract number 75D30123C1590. Any opinions, findings, and conclusions or recommendations expressed in this material are those of the author(s) and do not necessarily reflect the views of the United States of America Department of Health and Human Services, Centers for Disease Control and Prevention.

The authors thank the Delphi research group, members of InsightNet, and Lauren White and Tomas Leon of the California Department of Public Health for their helpful comments and suggestions.

\section{Code Availability}

All code used for these analyses can be found at \url{https://github.com/gthivierge/spatial-flu-forecasting} 

\bibliographystyle{unsrt}  
\bibliography{biblio}  

\section{Supplemental Figures}

\renewcommand{\thefigure}{S\arabic{figure}}
\setcounter{figure}{0}

\begin{figure}[htp]
    \centering
    \includegraphics[width=16cm]{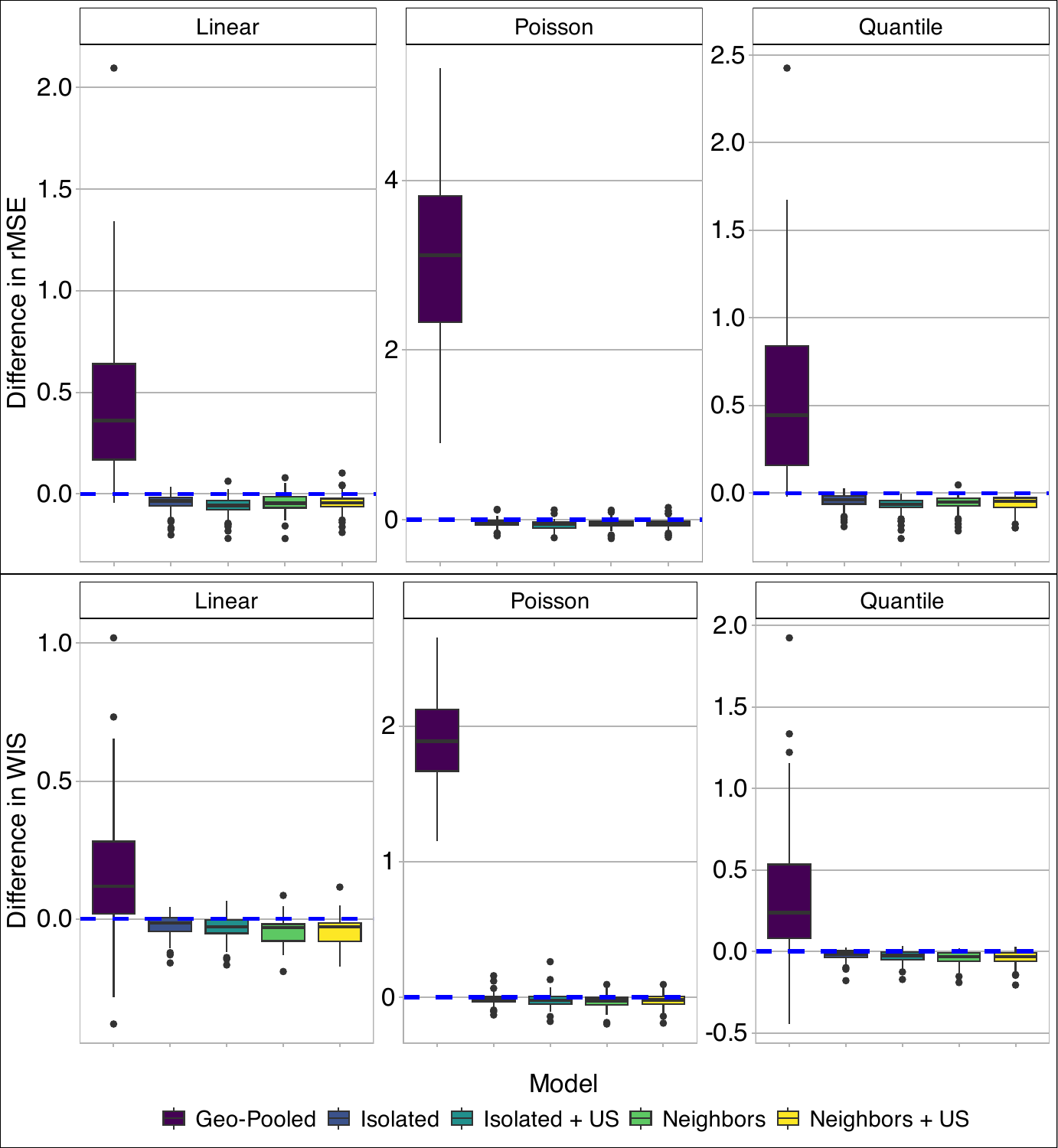}
    \caption{Autoregressive model variants demonstrated higher prediction accuracy than a last-value-carried-forward (LVCF) model, which in turn showed greater accuracy than the Geo-pooled model. The differences in rMSE and WIS are relative to the LVCF model (dashed line).}
    \label{fig:lvcf_baseline}
\end{figure}

\begin{figure}[htp]
    \centering
    \includegraphics[width=16cm]{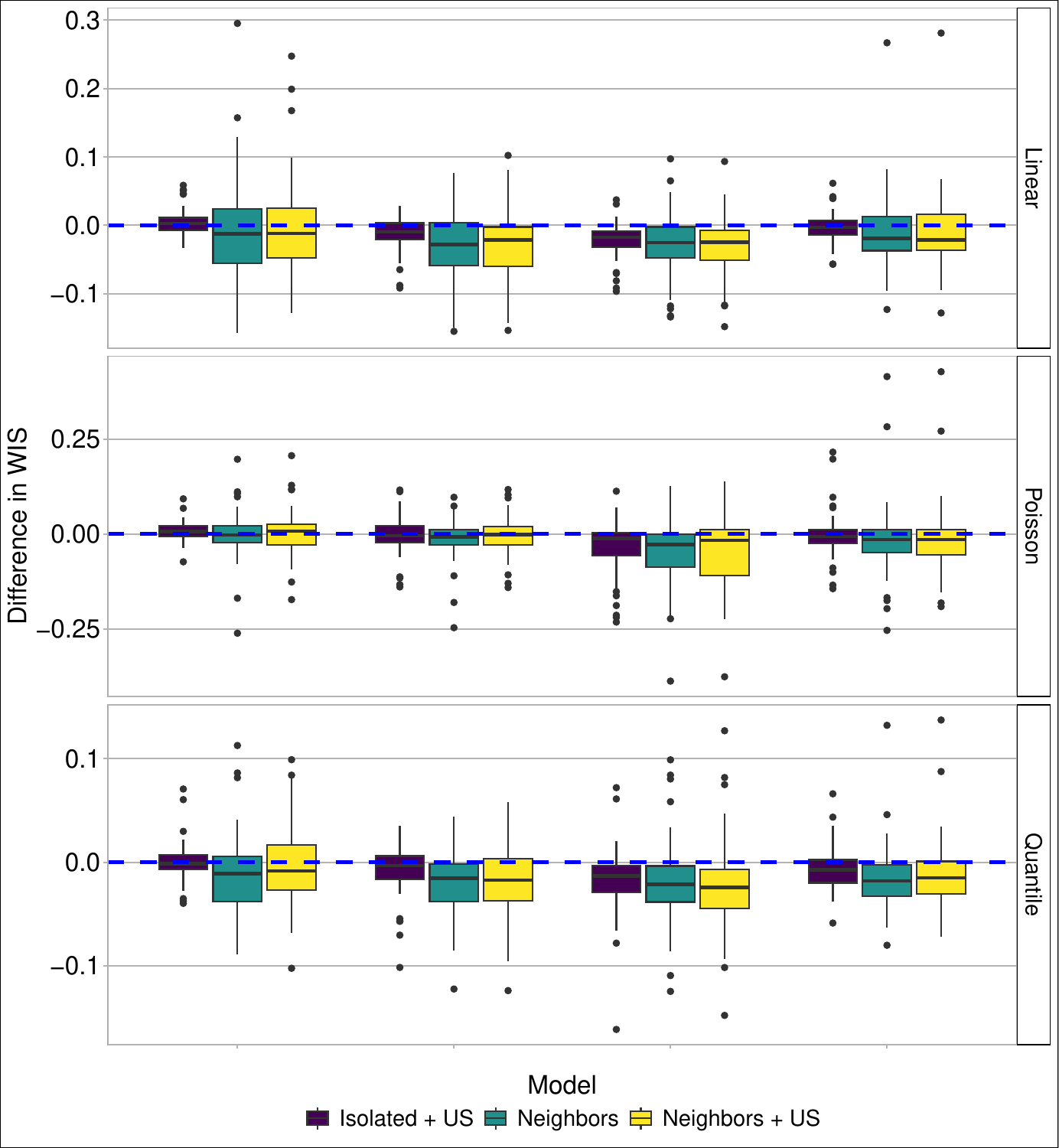}
    \caption{The relative benefit of spatial information for interval forecasting is not constant over time. When averaging WIS within states, the difference in accuracy between spatial models and the Isolated model varies across seasons.}
    \label{fig:isol_baseline_wis_szn}
\end{figure}

\begin{figure}[htp]
    \centering
    \includegraphics[width=16cm]{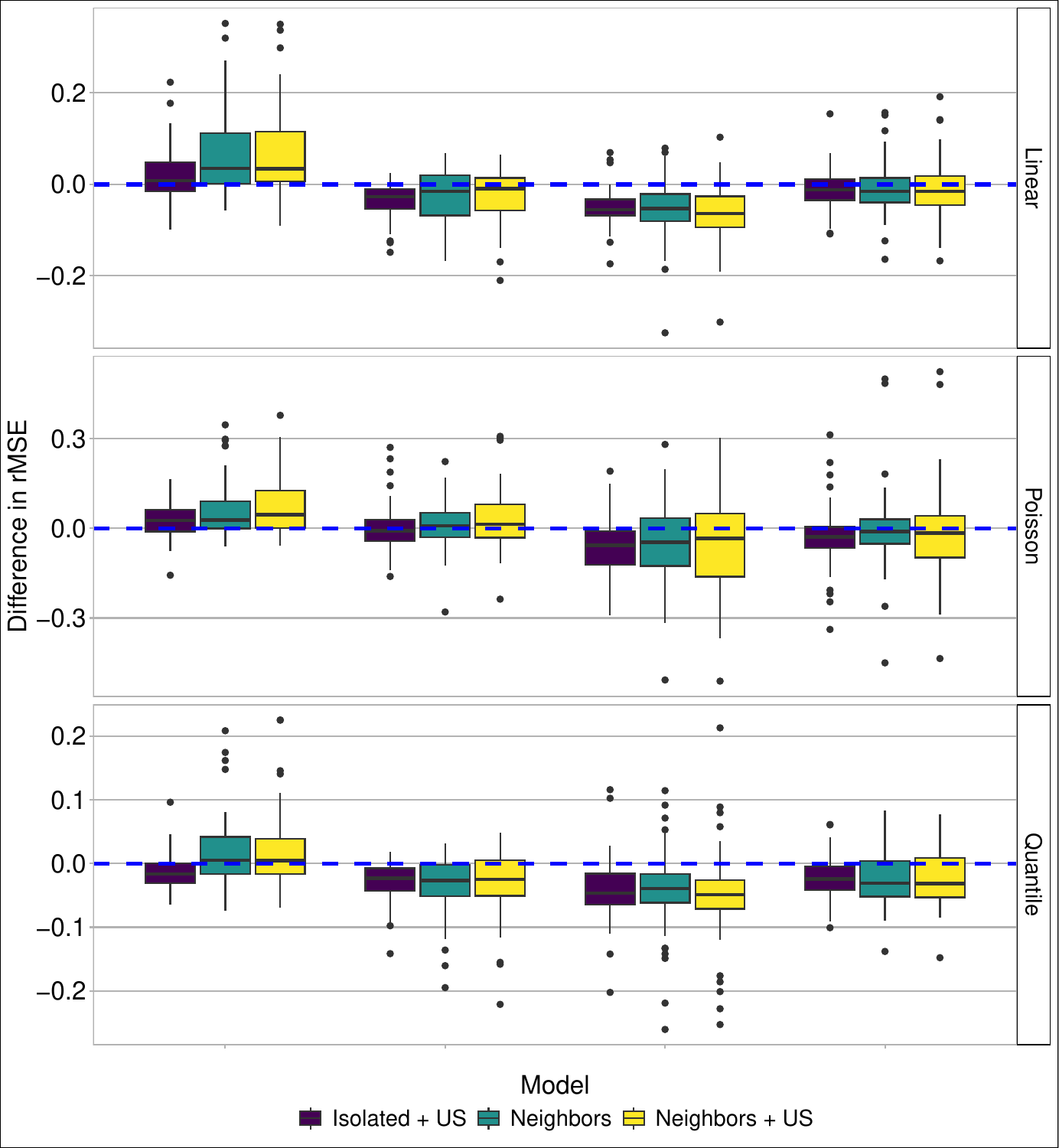}
    \caption{The relative benefit of spatial information for point predictions is not constant over time. When averaging rMSE within states, the difference in accuracy between spatial models and the Isolated model varies across seasons.}
    \label{fig:isol_baseline_rmse_szn}
\end{figure}


\begin{figure}[htp]
    \centering
    \includegraphics[width=18cm]{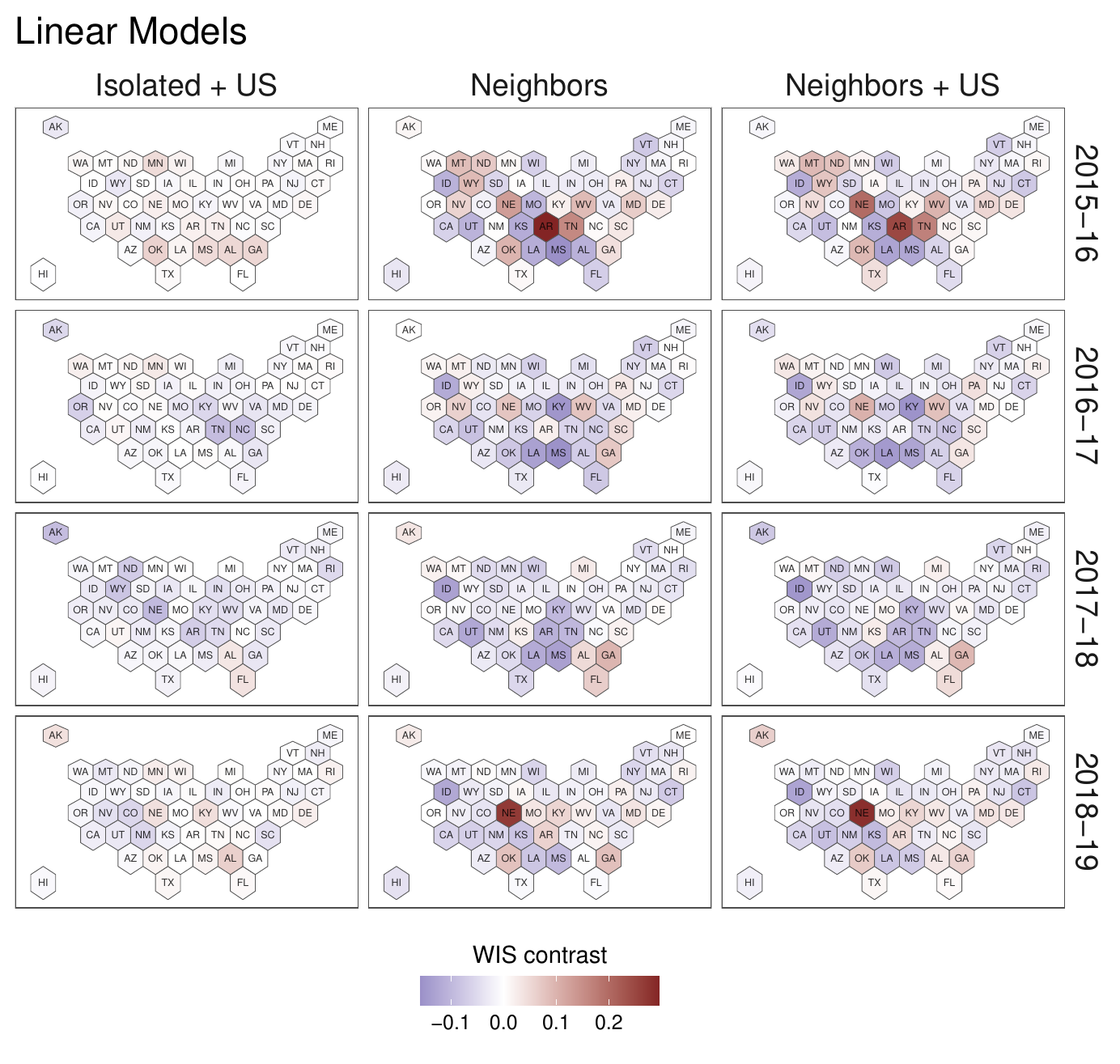}
    \caption{Difference in mean WIS relative to Isolated model by state, averaged by season, for linear models. At each time point, the WIS values for the predictions in each state were averaged and subtracted from the that of the Isolated model at the same time point. Negative values indicate that the model variant reduced WIS relative to the Isolated variant.}
    \label{fig:isol_bl_map_wis_szn_lin}
\end{figure}

\begin{figure}[htp]
    \centering
    \includegraphics[width=18cm]{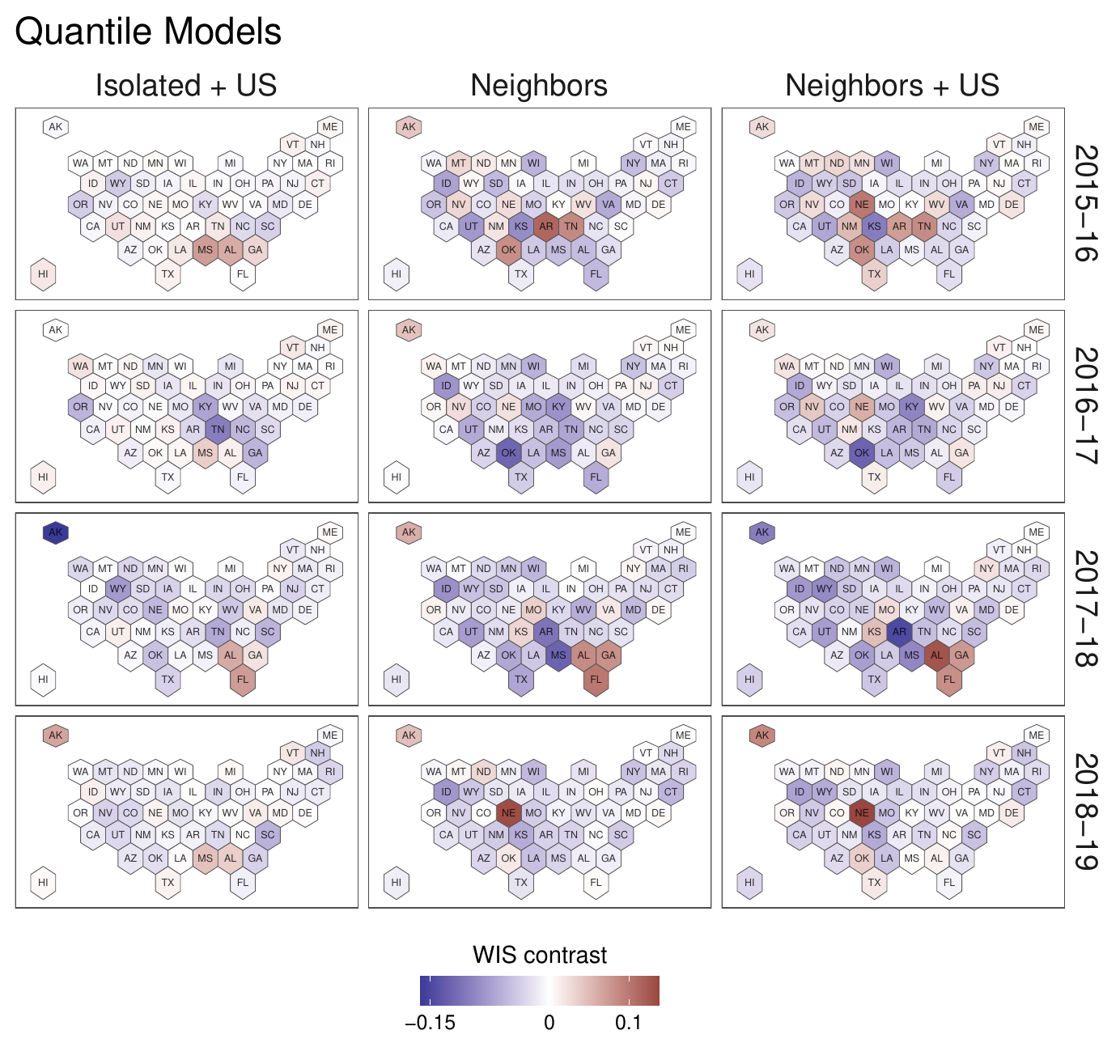}
    \caption{Difference in mean WIS relative to Isolated model by state, averaged by season, for quantile regression models. At each time point, the WIS values for the predictions in each state were averaged and subtracted from the that of the Isolated variant at the same time point. Negative values indicate that the model variant reduced WIS relative to the Isolated variant.}
    \label{fig:isol_bl_map_wis_szn_q}
\end{figure}

\begin{figure}[htp]
    \centering
    \includegraphics[width=18cm]{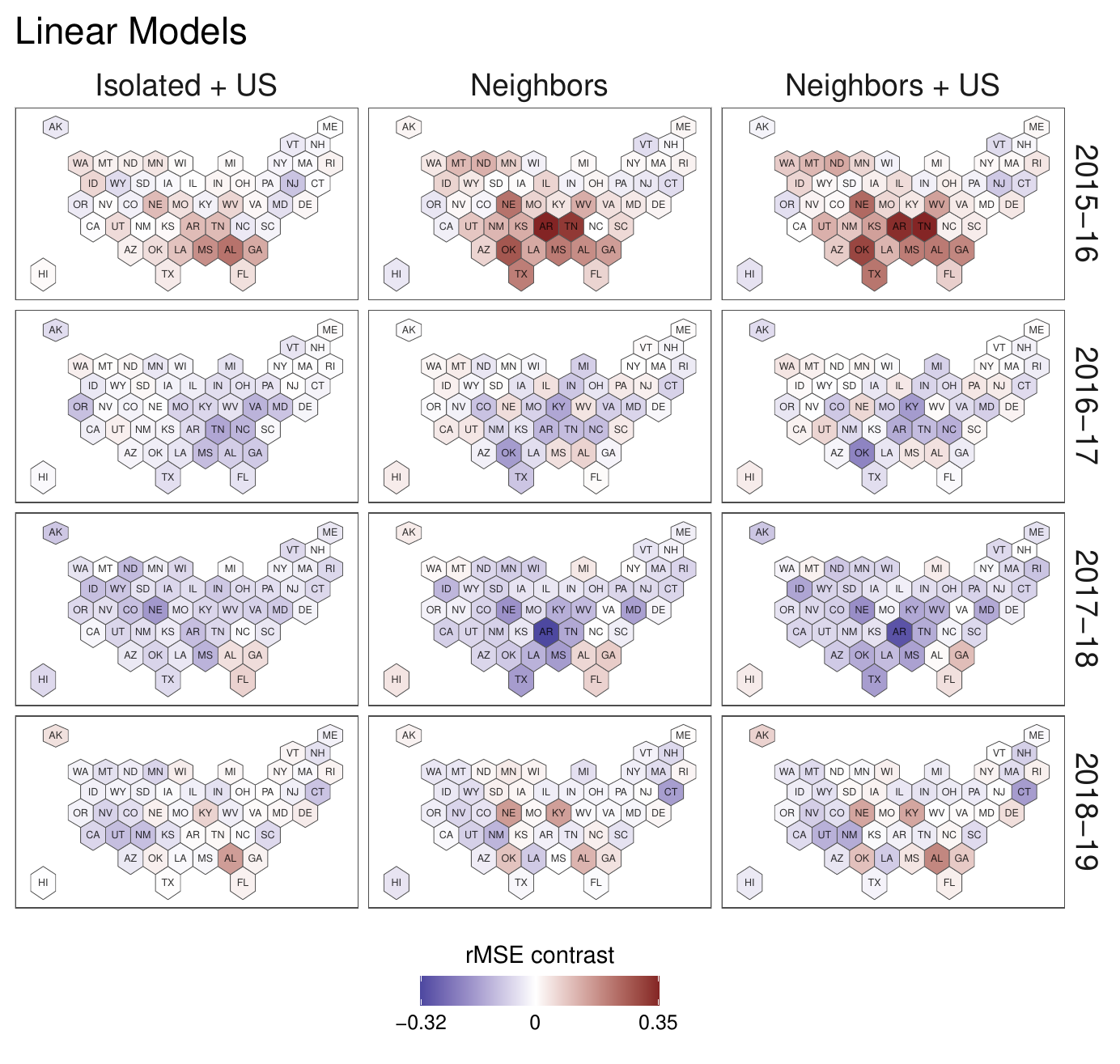}
    \caption{Difference in mean rMSE relative to Isolated model by state, averaged by season, for linear regression models. At each time point, the rMSE values for the predictions in each state were averaged and subtracted from the that of the Isolated model at the same time point. Negative values indicate that the model variant reduced rMSE relative to the Isolated variant.}
    \label{fig:isol_bl_map_rmse_szn_lin}
\end{figure}

\begin{figure}[htp]
    \centering
    \includegraphics[width=18cm]{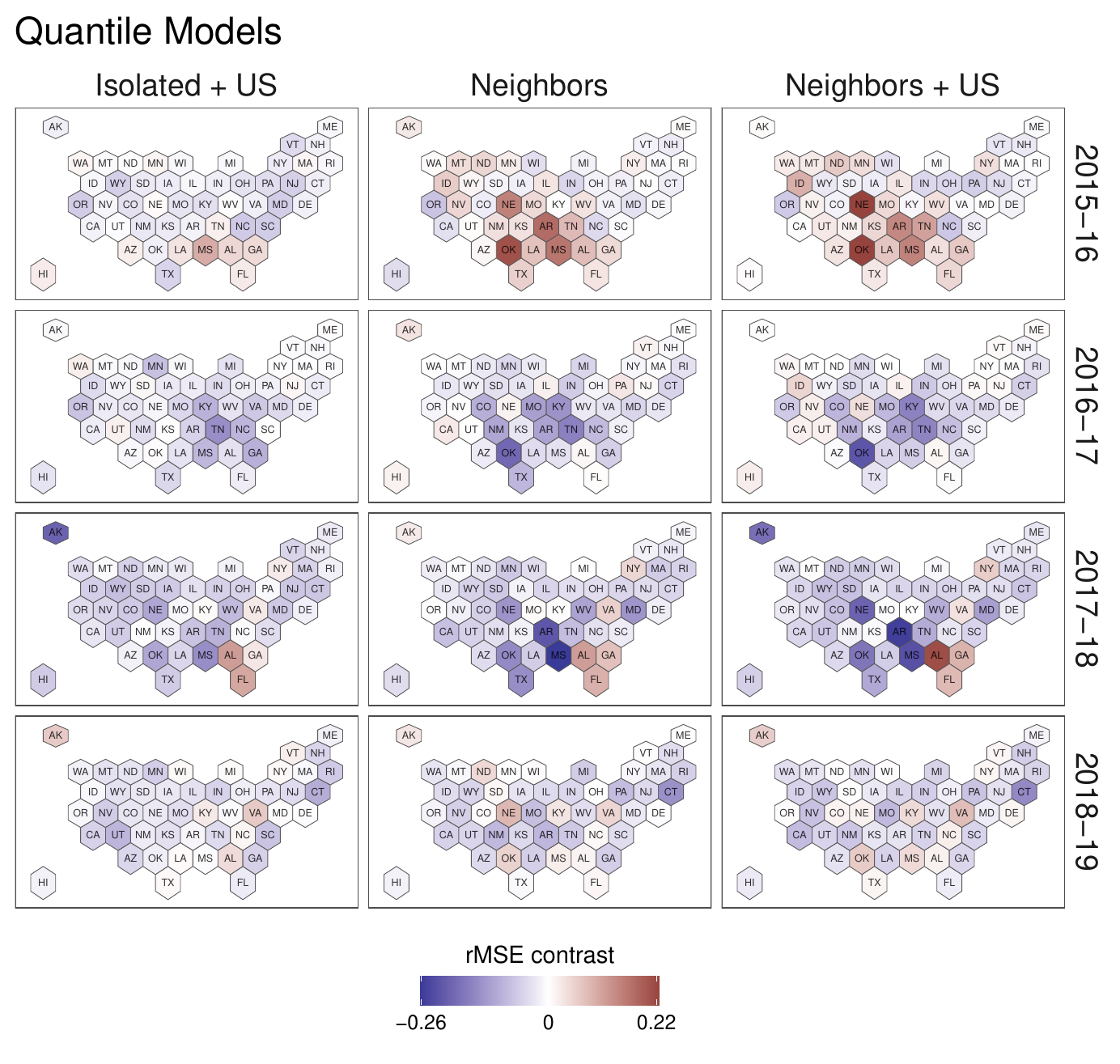}
    \caption{Difference in mean rMSE relative to Isolated model by state, averaged by season, for quantile regression models. At each time point, the rMSE values for the predictions in each state were averaged and subtracted from the that of the Isolated model at the same time point. Negative values indicate that the model variant reduced rMSE relative to the Isolated variant.}
    \label{fig:isol_bl_map_rmse_szn_q}
\end{figure}

\end{document}